\documentstyle[amssymb,preprint,aps]{revtex}
\newtheorem{theorem}{Theorem}
\newtheorem{acknowledgement}[theorem]{Acknowledgement}

\begin{document}
\title{Superconducting properties of $\left[ BaCuO_{x}\right] _{2}/\left[ CaCuO_{2}%
\right] _{n}$ artificial structures with ultrathick $CaCuO_{2}$ blocks}
\author{G. Balestrino, S. Lavanga$^{\ast }$, P. G. Medaglia, S. Martellucci, A.
Paoletti, G. Pasquini, G. Petrocelli, A. Tebano, and A. A. Varlamov}
\address{INFM-Unit\`{a} di Ricerca ''Tor Vergata'', Dipartimento di Scienze e\\
Tecnologie Fisiche ed Energetiche,\\
Universit\`{a} di Roma ''Tor Vergata'', Via di Tor Vergata 110, 00133 Roma}
\author{L. Maritato, M. Salvato}
\address{INFM-Unit\`{a} di Ricerca di Salerno, Via Allende 5, 84081 Salerno}
\date{\today }
\maketitle

\begin{abstract}
The electrical transport properties of $[BaCuO_{x}]_{2}/[CaCuO_{2}]_{n}$ $%
(CBCCO-2\times n)$\ underdoped high temperature superconducting
superlattices grown by Pulsed Laser Deposition have been investigated.
Starting from the optimally doped $CBCCO-2\times 2$ superlattice, having
three $CuO_{2}$ planes and $T_{c}$ around $80$ $K,$ we have systematically
increased the number $n$ up to $15$ moving toward the underdoped region and
hence decreasing $T_{c}$. For $n>11$ \ the artificial structures are no
longer superconducting, as expected, for a uniformly distributed charge
carriers density inside the conducting block layer. The sheet resistance of
such artificial structures ($n\approx 11$) turns out to be quite temperature
independent and close to the 2D quantum resistance $26$ $k\Omega $. A
further increase of the number of $CuO_{2}$ planes results in an
insulator-type dependence of $R(T)$ in the wide range of temperatures from
room temperature to $1$ $K$. The value of the sheet resistance separating
the Superconducting and the Insulating regimes supports the fermionic
scenario of the Superconductor-Insulator transition in these systems.
\end{abstract}

\subsection{Introduction}

Recently\ a large effort has been devoted to the layer-by-layer growth of
high temperature superconducting cuprate (HTS) superlattices. Among them,
particularly interesting are the HTS superlattices $%
[BaCuO_{x}]_{m}/[CaCuO_{2}]_{n}$ $(CBCCO-m\times n)$, with $m\simeq 2$\cite
{BAL APL97}. These artificial structures are grown by Pulsed Laser
Deposition (PLD) stacking in sequence nonsuperconducting individual layers
of $BaCuO_{x}$ and $CaCuO_{2}$ In the $CBCCO-m\times n$ unit cell, the
superconducting block, named Infinite Layer (IL) block, consists of $n$
epitaxial layers of $CaCuO_{2}$, while the Charge Reservoir (CR) block is
made of $\ m$ epitaxial layers of $BaCuO_{x}$\cite{BAL PhysicaC97}. In the
case of $[BaCuO_{x}]_{2}/[CaCuO_{2}]_{2}$ ($2\times 2$ superlattices) grown
at relatively high molecular\ oxygen pressure ($\simeq $ $1$ $mbar$) and at
a temperature of about $650$ $%
%TCIMACRO{\UNICODE[m]{0xb0}}%
%BeginExpansion
{{}^\circ}%
%EndExpansion
C$, it has been shown that the doping is nearly optimal $(p\simeq 0.18-0.19$
holes$/CuO_{2}$ plane) and $T_{c}$ $\simeq 80$ $K$ \cite{BAL Solid
state98,BAL PRB99}.

In our previous work on the $CBCCO-2\times n$ system, we investigated both
the dependence of the critical temperature on $n$ (for $n$ ranging from 1 to
6) and the influence of the interfacial structural disorder, typical for
these artificial structures, on\ the electrical transport properties \cite
{BAL PRB99,BAL PRB98}. It was found that the critical temperature reaches
the maximum value of about $80$ $K$ for $n$ between 2 and 3. It decreases
both for $n$ larger than 3 and smaller than 2. Such an effect was explained
on the basis of the structural disorder (especially effective for small $n$)
and of the variation of the effective carriers concentration with $n$.
Furthermore it was demonstrated that some peculiar features of the
temperature dependence of the resistivity (its relatively high residual
value, its pronounced rounding above $T_{c}$ and its saturation above room
temperature) can be well explained by taking into account the role of the
interfacial structural disorder that arises at the blocks interfaces \cite
{BAL PRB99}. Following this considerations, we believe that the role of the
structural disorder is strongly reduced for the superlattices with thick IL
blocks, negligible for the thickest ones.

In this paper we report on the investigation of the electrical transport
properties of $[BaCuO_{x}]_{2}/[CaCuO_{2}]_{n}$ superlattices having
ultrathick IL\ layers. For this purpose we have grown many $\ 2\times n$
superlattices having the same CR block, consisting of 2 epitaxial layers of $%
BaCuO_{x}$, but different superconducting blocks containing $(n+1)\ CuO_{2}$
planes \cite{BAL PRB98}. We varied the number of $CaCuO_{2}$ layers per cell
from 2 up to 15, keeping constant all the growth parameters and the
thickness of the CR block of the superlattices. In this way we decreased the
average doping level per $CuO_{2}$ plane inside the superconducting block,
starting from the optimally doped $CBCCO-2\times 2$ superlattice with $T_{c}$
around $80$ $K$. An important experimental finding was that for $n>11$
(corresponding to the critical value $p\simeq 0.04$ holes$/CuO_{2}$ plane)
the critical temperature goes to zero and the resistivity versus temperature
follows closely the behavior predicted by the variable range hopping
mechanism. Such a result indicates that the distribution of the charge
carriers in the IL block is probably quite uniform without sizeable
confinement at the interfaces between the IL block and the CR parent block.
On the contrary, in the case of the charge localization at the CR/IL
interfaces, one would expect a saturation of the critical temperature value
for a sufficiently thick IL block (as in the case of twin-planes
superconductivity \cite{BKh80}).

Therefore, starting from the optimally doped $CBCCO-2\times 2$ superlattice
we have systematically increased the number $n$ of $CuO_{2}$ planes per
cell, moving toward the underdoped part of the phase diagram and hence
decreasing $T_{c}$. In this way the doping level (per $CuO_{2}$ plane)
becomes smaller and smaller, until a critical value of the IL block
thickness is reached at which superconductivity disappears. We show that,
for the $2\times 11$ superlattice, the superconducting critical temperature
tends to zero while $R_{_{\square }}$ approaches the value of $25.8$ $%
k\Omega $ close to the universal quantum resistance for the $2D$
Superconductor-Insulator (S-I) transition \cite{Gur PRB 96}. For $n>11$ the $%
R(T)$ behavior was found to be of the insulator-type. The results obtained
suggest that the S-I transition, occurring at $n\simeq 11,$ can be explained
within the fermionic scenario, where the insulating state of $2D$ electrons
system is reached when the charge carriers density is so low that$\
k_{F}l\simeq 1.\ $

\subsection{Experimental results}

Artificial structures with $n$ ranging from $2$ to $15$ have been grown by
PLD on $SrTiO_{3}(001)$ substrates. Two targets, with $CaCuO_{2}$ and $%
BaCuO_{2}$ nominal composition respectively, mounted on a multitarget
system, were used. The oxygen pressure and the substrate temperature, during
the deposition, were $1$ $mbar$ and $650$ $%
%TCIMACRO{\UNICODE[m]{0xb0}}%
%BeginExpansion
{{}^\circ}%
%EndExpansion
C$ respectively. Further details on the\ growth procedure are given in Ref. 
\cite{BAL PhysicaC97}. Films have thickness ranging from $600$ to $1000$ \AA %
. The samples crystallographic structure were characterized by X-ray
diffraction (XRD) analysis.

The unit cells thickness $\Lambda $ varies between $15.2$ \AA $\ $($2\times 2
$ superlattices) and $56.8$ \AA\ ($2\times 15$ superlattices). The thickness
of the IL block has been varied increasing the number of laser shots on the $%
CaCuO_{2}$ target. All films were found to have a very small value of the
mosaic spread ($\simeq 0.08%
%TCIMACRO{\UNICODE[m]{0xb0}}%
%BeginExpansion
{{}^\circ}%
%EndExpansion
$), close to the substrate one ($\simeq 0.06%
%TCIMACRO{\UNICODE[m]{0xb0}}%
%BeginExpansion
{{}^\circ}%
%EndExpansion
$). In Fig.2 the $SL_{0}(002)$ rocking curve of the superconducting $2\times
2$ superlattice (a) and the $SL_{+1}(001)$ of the insulating $1.9\times 14$
superlattice (b) are shown, the superlattices\ peaks, $SL_{\pm N}$ $(00l)$,
are the $N^{th}$\ order satellite peaks of the substrate $00l$ reflections,
the $zero^{th}$ order being the average structure peak of the artificial
structure.

The XRD spectra of the $[BaCuO_{x}]_{2}/[CaCuO_{2}]_{n}$ superlattices are
shown by empty dots in Fig.1 for four different values of $n$. The effective
thickness of both the CR and IL block were deduced by a kinematical
simulation of the diffraction spectra (full lines in Fig.1). This simulation
is achieved considering that a two dimensional layer-by-layer growth occurs
and that layers of mixed composition are corrugated to adjust the internal
stresses. A random disorder is added to take into account the experimental
dispersion in the amount of deposited material in each iteration \cite{IN
Press}. In our simulations the random thickness fluctuations for each
deposited layer is found to be indipendent from $n$. Furthermore a small
decrease in the deposition rate during the growth has been considered in
order to reproduce the drastic suppression of the $SL_{\pm 3}$ and $SL_{\pm
4}$ peaks. Such an effect is expected due to the progressive\ damage of the
target surface during the growth. The\ agreement between the experimental\
data and the simulated spectra is very good. All the major features of the
diffraction spectra are reproduced by the simulations. 

Preliminary asymmetrical reflections XRD measurements, carried out at the
European Synchrotron Radiation Facility, have shown that the in-plane
crystallographic directions $a$, $b$ of the superlattices are aligned with
those of the $SrTiO_{3}(001)$ substrates with an uncertainty of $0.15%
%TCIMACRO{\UNICODE[m]{0xb0}}%
%BeginExpansion
{{}^\circ}%
%EndExpansion
$ \cite{Aruta}.

Resistivity measurements were performed by the standard four-probe dc
technique. Contacts were made by silver epoxy directly on the substrate
before the film deposition, in order to avoid any chemical reaction between
the superlattices and the solvent utilized in the silver epoxy. The probe
current density was about $100$ $A/cm^{2}$. The temperature was varied at a
typical rate of $1K/min$, with an accuracy of $10$ $mK.$ The sheet
resistance $R_{_{\square }}$ was calculated, for each superlattice, assuming
the distance between two $2D$ successive conducting block layers to be equal
to the specific modulation length $\Lambda $ : $R_{_{\square }}=%
%TCIMACRO{\dfrac{\rho _{ab}}{\Lambda }}%
%BeginExpansion
{\displaystyle{\rho _{ab} \over \Lambda }}%
%EndExpansion
.$

The behavior of sheet resistance versus temperature is shown in Fig.3a for
several superlattices with different $n$ values. By increasing the number of 
$CuO_{2}$ planes in the IL block, the following striking features can be
noticed:

a). $T_{c}$ goes smoothly to zero. For the $2\times 11$ superlattices
(having $12$ $CuO_{2}$ planes in the IL block) the critical temperature
(zero resistance point) is $1$ $K$, while the $2\times 14$ and $2\times 15$
superlattices show an insulating behavior with no trace of superconductivity
above $1$ $K.$

b). The sheet resistance $R_{_{\square }}$ becomes larger.\ All the $CBCCO$
films with $n>8$\ show an upturn of resistivity with a negative $d\rho /dT$\
(see the Fig.3b) at low temperature above the $\sup $erconducting transition.

c). The crossover between metallic (positive $d\rho /dT$ ) and insulating
(negative $d\rho /dT$ ) low temperature behavior occurs for $n>8$. Films
with the thickest IL block have sheet resistance higher than $30$ $k\Omega .$

Curves belonging to {\em superconducting} and {\em insulating} phases could
be ideally separated by a separatrix lying near the $2\times 11$
superlattice $R(T)$ dependence. This ideal separatrix curve has a sheet
resistance of the order of quantum $2D$ resistance $(\simeq 26$ $k\Omega )$
(see the detailed discussion of the properties of this critical region in 
\cite{AMP00}). As one can see from Fig. 5 films with $n>11$ are not
superconducting: this behavior was expected because we can estimate a doping
level $p<0.04$ holes$/CuO_{2}$ plane, below the critical value for
superconductivity in many HTS cuprates \cite{Tinkham,Tak PRL92}. All films
with $15$ or more $CuO_{2}$ planes in the IL block showed the variable range
hopping (VRH) behavior typical of 2D insulators. This point is stressed in
the inset of Fig.5, where the $ln(R)$ is reported as a function of $%
(1/T)^{1/3}$. The excellent linear behavior of the experimental data can be
noted. The above scenario is also enforced by the inset of Fig.4: here a
clear inverse correlation between the critical temperature and the sheet
resistance can be noted: $T_{c}$ goes to zero as $R_{_{\square }}$ reaches $%
25-30$ $k\Omega $.

\subsection{Discussion}

The first important finding concerns the charge carriers distribution over
the IL block. The disappearance of superconductivity for the samples with
thick enough IL blocks indicates that the carriers are distributed quite
homogeneously along the $c-$axis and that their density decreases increasing 
$n$. Assuming $\rho (n)=\rho _{opt}$ $\frac{3}{n+1}$,\ with $\rho
_{opt}\simeq 0.18$, one can estimate the critical doping value $\rho (n=11)$%
\ $\simeq 0.045$ at which the phenomenon of superconductivity disappears. It
is clear that this value is very close to the critical doping of the HTS
materials ($\simeq 0.04$). \cite{Tinkham,Tak PRL92}

Another interesting aspect of our investigation is the new way to drive the
system through the S-I transition: namely, leaving unchanged the overall
number of carriers provided by the CR block, we have varied the effective
carriers density, varying the number of\ ''conducting'' $CuO_{2}$ planes in
the IL block.

The S-I transition has been widely studied during the last years in a
variety of disordered and underdoped systems: in granular thin metallic film
such as $Bi,$ $Sn,$ $Pb,$ $Al$ \cite{Phys Today}, and in high-$T_{c}$
cuprates such as irradiated $Y-Ba-Cu-O$ and $Bi-Sr-Ca-Cu-O$ \cite{Gur PRB
96,Rullier Phys C95}, $Zn-$doped $La-Ba-Ca-Cu-O$ \cite{Singh PRB97}, $Y-$%
doped $Bi-Sr-Ca-Cu-O$ \cite{MandrusPRB91}, $Ce-$doped $Nd-Cu-O$ \cite{Tanda
PRB91}, oxygen-deficient $Y-Ba-Cu-O$ single crystals \cite{Veal PRB92}, $Pr-$%
doped or $Zn-$doped $Y-Ba-Cu-O$ \cite{Levin PRL98,chien PRL91}, ultrathin $%
Dy-Ba-Cu-O$ films \cite{Gold PRB91}. In this context it is evident that our
superlattices are not the simplest objects for the study of such fundamental
phenomenon like the S-I transition. Nevertheless they show some advantages
relative to other systems. Namely, due to the complexity of the unit cells
of the HTS materials, an evident ambiguity exists concerning the definition
of the effective thickness $t$ of the conducting layer necessary to
calculate $R_{_{\square }}.$ It can be seen that, in the case of many HTS
compounds, the specific choice of the thickness of the conducting layer
(either the whole unit cell that is the CR plus the IL block, or the IL
block alone, or the single $CuO_{2}$ plane) can result in a large
uncertainty in the $R_{_{\square }}$ value. For instance, in the case of $%
Bi_{2}Sr_{2}Y_{x}Ca_{1-x}Cu_{2}O_{8}$ compound \cite{MandrusPRB91} the
thickness $t$\ of the conducting layer has been chosen as the $%
CuO_{2}/Ca/CuO_{2}$ layers\ distance (namely the half of $c-$axis lattice
constant, $t\simeq 15$ \AA ), instead in Ref. \cite{Veal PRB92} $t$ is equal
to the $c-$axis lattice constant of $Y_{1}Ba_{2}Cu_{3}O_{7-x}$ ($t\simeq
11.8 $ \AA ). On the other hand, in the case of $Nd_{2-x}Ce_{x}Cu_{1}O_{4}$
single crystal \cite{Tanda PRB91}, $t$ is the lattice spacing between $%
CuO_{2}$ layers ($t\simeq 6$ \AA ). This ambiguity turns out to be crucial
for the possibility of judging which type of scenario of the 2D S-I
transition (bosonic or fermionic) is realized in practice \cite{Larkin99}.

The first mechanism (bosonic), is based on the duality hypothesis. In fact,
both classical and quantum fluctuations reduce the superfluid density $\rho
_{s}$ and therefore, suppress the Berezinski-Thouless-Kosterlitz transition
temperature $T_{c}^{{\rm BKT}}$. The magnitude of this suppression is
determined by the Ginzburg-Levanyuk number $Gi_{2}$ which turns out to be of
the order of 1 when the dimensionless conductance $g=$ $%
%TCIMACRO{\dfrac{\sigma }{e^{2}/h}}%
%BeginExpansion
{\displaystyle{\sigma  \over e^{2}/h}}%
%EndExpansion
\rightarrow g_{c}\sim 1.$ As a result, at this level of disorder, the
superfluid density, simultaneously with $T_{c}^{{\rm BKT}},$ becomes zero.
In the vicinity of this critical carriers concentration is $T_{c}^{{\rm BKT}%
}\ll T_{c(0)}$ (where $T_{c(0)}$ is the mean field critical temperature) and
in the temperature range $T_{c}^{{\rm BKT}}\ll T\ll T_{c(0)}$, the
conductivity is realized by fluctuating Cooper pairs. The problem of quantum
liquid motion near the quantum phase transition can be approached from a
second \ view point. It can be said that, with the increase of $Gi_{2}$, the
role of quantum fluctuations grows and that fluctuating vortices, carrying
the magnetic flux quantum $\Phi _{0}=h/2e$, are generated. The duality
hypothesis \cite{FFH91} assumes that, at the critical point, the pair and
vortex liquid density flows are equal, giving, for the conductivity the
universal value, at the critical point: $\sigma _{c}=\frac{2e}{\Phi _{0}}=%
\frac{4e^{2}}{h}.$

The second, fermionic, mechanism of the S-I transition is related to the
renormalization of the inter-electron interaction in the Cooper channel by
the long-range Coulomb repulsion, specific to dirty $2D$ superconductors 
\cite{o,fu}.\ As long as the correction to the unrenormalized BCS transition
temperature $T_{c(0)}$ is still small, it is found that: $%
T_{c}=T_{c(0)}\left( 1-%
%TCIMACRO{\dfrac{1}{12\pi ^{2}g}}%
%BeginExpansion
{\displaystyle{1 \over 12\pi ^{2}g}}%
%EndExpansion
\ln ^{3}%
%TCIMACRO{\dfrac{\hbar }{k_{B}T_{c(0)}\tau }}%
%BeginExpansion
{\displaystyle{\hbar  \over k_{B}T_{c(0)}\tau }}%
%EndExpansion
\right) .$ The suppression of $T_{c}$ down to zero in this case may occur
even for $g\gg 1.$ The renormalization group analysis gives \cite{fi}, for
the critical value of conductance: $g_{c}=\left( \frac{1}{2\pi }\ln \frac{%
\hbar }{k_{B}T_{c(0)}\tau }\right) ^{2}.$ At high degree of disorder (or low
level of doping), $T_{c(0)}\tau $ is small enough, so that $\ln \frac{\hbar 
}{k_{B}T_{c(0)}\tau }>5$: the fermionic mechanism of the critical
temperature suppression turns out to be of primary importance. In this case $%
g_{c}>2/\pi $: the boson mechanism turns out to be insignificant. On the
contrary, if $\ln \frac{\hbar }{k_{B}T_{c(0)}\tau }<4,$ the correction for $%
T_{c}$ is small even for $g_{c}=2/\pi $, so the boson mechanism becomes of
primary importance. The typical experimental values of $g_{c}$ are in the
region $g_{c}\sim 1$, and do not differ dramatically from the predictions of
the boson duality assumption $g_{c}=\frac{2}{\pi }$ (by estimating for the $%
2\times 2$ superlattices \cite{BAL PRB98} $\tau \simeq 2\times 10^{-15}$ $%
sec $, $T_{c(0)}=80$ $K,$ $g_{c}$ turns out to be quite larger than $%
%TCIMACRO{\dfrac{2}{\pi }}%
%BeginExpansion
{\displaystyle{2 \over \pi }}%
%EndExpansion
$)$.$ This is why the precise evaluation of the experimental value of the
sheet resistance is so important. In the case of our samples, with
ultrathick IL blocks, the ambiguity in determining the conducting layer
thickness is minimal. In fact, this ambiguity is of the order of the CR
block thickness in comparison with that one of the much thicker IL block.
Therefore the value of $R_{_{\square }}\simeq 25-30$ $k\Omega $, found
according to our experiments, is highly reliable and allows us to conclude
that the observed S-I transition has a fermionic character. An additional
argument in favor of{\it \ }this conclusion can be found in the logarithmic
growth of the resistance in the metal phase of the $2\times (9\div 11)$
samples (see Fig.5) which is\ coherent with the weak localization theory.

\begin{acknowledgement}
The authors (G. B. and A. A. V.) are grateful to B. L. Altshuler, A. Goldman
and A. I. Larkin for valuable discussions. The financial support of the
research project HTSS of INFM is acknowledged.
\end{acknowledgement}

\begin{figure}[tbp]
\caption{XRD spectra of four $m\times n$ superlattices with ultrathick IL
blocks (open circles) and the corresponding simulated spectra (full line)
obtained by the procedure described in Ref.\protect\cite{IN Press}. The $%
SrTiO_{3}(001)$ peaks are labeled by the asterisks.}
\end{figure}

\bigskip

\begin{figure}[tbp]
\caption{Rocking curves: a) $SL_{0}(002)$ peak of the $2\times 2$
superconducting superlattice. b) $SL_{+1}(001)$ peak of the $1.9\times 14$
insulating superlattice. The full width at half maximum (FWHM) were
estimated by Lorentzian curve fits.}
\end{figure}

\begin{figure}[tbp]
\caption{a) Sheet resistance versus temperature of $CBCCO-m\times n$
superlattices in which $m$ is nearly 2. b) First derivative of the
resistance, normalized to its $300$ $K$ value, of the same superlattices.}
\end{figure}

\begin{figure}[tbp]
\caption{Critical temperature (zero resistance point) versus the number of $%
CuO_{2}$ planes in the IL block. The vertical dotted line ideally divides
the overdoped from the underdoped region. Inset: critical temperature versus
the sheet resistance. Here the sheet resistance values were obtained as the $%
T=0$ $K$ linear extrapolation of $\protect\rho _{ab}/\Lambda $, where $%
\Lambda $ is the thickness of the unit cell.}
\end{figure}

\begin{figure}[tbp]
\caption{Normalized resistance versus $T$\ ($20$ $K<T<300$ $K$) in
semilogarithmic scale of two ultrathick superconducting superlattices.
Inset: $2D$\ VRH behavior of the thickest structures.}
\end{figure}

\end{document}